\documentstyle[aps,prl,twocolumn,epsfig]{revtex}
\begin{document}
\draft
\title{A Systematic Formulation of Slow Polaritons in Atomic Gases}
\author{G. Juzeli\={u}nas $^{1,2}$ and H. J. Carmichael $^{2}$}
\address{$^{1}$ Institute of Theoretical Physics and Astronomy, A. Go\v{s}tauto 12, 
Vilnius 2600, Lithuania\\
$^{2}$ Department of Physics, University of Oregon, Eugene, OR 97403-1274, USA}
\date{\today}
\maketitle

\begin{abstract}
We formulate a theory of slow polaritons in atomic gases and apply it to the slowing
down, storing, and redirecting of laser pulses in an EIT medium. The normal modes
of the coupled matter and radiation are determined through a full diagonalization of
the dissipationless Hamiltonian. Away from the EIT resonance where the polaritons
acquire an excited-state contribution, lifetimes are introduced as a secondary step.
With detuning included various four-wave mixing possibilities are analyzed. We investigate
specifically the possibility of reverting a stopped polariton by reversing the control
beam. 
\end{abstract}

Recently, electromagnetically induced transparency (EIT)
\cite{Arimondo:96,Harris:97,Scully:book,Bergman:98} was shown to slow down
dramatically\cite{Hau:99}, or even to stop completely \cite{Hau:01,Phill:01},
laser pulses in atomic gases. The experiments involve media of three-level atoms
interacting with two lasers---a control beam and a probe pulse. The atoms have two
hyperfine ground states, $\left|g\right\rangle$ and $\left|q\right\rangle$, and
an electronically excited state $\left|e\right\rangle$, as illustrated in Fig.~1(a).
Level $g$ is populated initially, before applying the probe pulse to couple
$g$ and $e$. The role of the control beam is to introduce a transparency window
so that the probe pulse propagates slowly in the medium.
   
Such behavior can be understood in terms of a branch of slow polaritons appearing
between two close atomic resonances, as considered by Juzeli\={u}nas
(see Fig.~2(b) in \cite{Juz:96}).  Indeed, the control laser couples the states
$\left|q\right\rangle$ and $\left|e\right\rangle$ dynamically, bringing level~$q$
into resonance with the excited level~$e$. The excited level splits, then, into
the doublet shown in Fig.~1(b), giving precisely the level structure required to
form a branch of slow polaritons.

\begin{figure}[hb] 
\leavevmode
\centering
\epsfxsize=3.0in
\epsffile{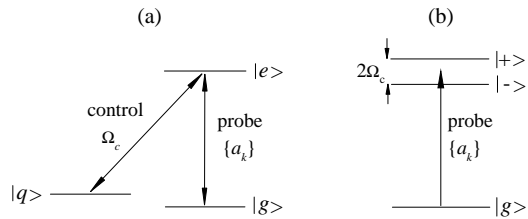}
\vskip0.1in
\caption{(a) Atomic level scheme in the $\Lambda$ configuration of EIT.
(b) Equivalent scheme incorporating the dynamical coupling of $q$ and $e$
characterized by the Rabi frequency $\Omega_c$.
}
\end{figure}

Polaritons are the normal modes of a combined system of radiation and matter
and are a familiar subject in solid state physics. Over the last decade the
polariton idea has been applied widely to describe the quantized radiation
field in dielectric media
 \cite{Juz:96,Knoe:89,Svist:90,Politz:91,Barn:92,Drum:99,Juz:00}. Most studies,
however, considered media of two-level atoms, and hence cannot accommodate
the slow EIT polaritons. Slow polaritons appear in the analysis beyond two levels
\cite{Juz:96,Drum:99,Juz:00}; although the existing theoretical work does not
deal with EIT, specifically. EIT (dark state) polaritons were first considered
theoretically by Mazets and Matisov \cite{Mazets:96}, and later by Fleischhauer
and Lukin\cite{Fleis:00,Fleis:01} who suggested storing the probe pulse
(stopping the polariton) by adiabatically switching off the control laser.

In this paper we present a systematic description of slow polaritons in EIT media.
The theory is developed for atomic Bose-Einstein condensates (BECs), but is also
applicable to ordinary atomic gases. In contrast to previous work
\cite{Mazets:96,Fleis:00,Fleis:01}, we explicitly diagonalize the full Hamiltonian,
including detuning from the EIT resonance and the contact interaction, and without
making the rotating wave approximation in the interaction with the probe field.
Away from the EIT resonance the polaritons acquire an excited-state contribution
which leads to a finite radiative lifetime. With detuning included various four-wave
mixing possibilities are suggested and we specifically propose a scheme for reverting
a stopped polariton. We also apply our formulation to the recent experiments
\cite{Hau:01,Phill:01} where it is apparent from the explicitly constructed
polariton modes that the storage and retrieval of the probe pulse takes place at the
medium boundaries (to within an extremely small correction) and stopping the polariton
can be achieved even by a sudden ``turn-off'' of the control laser.

Consider a gas of Bose atoms with an internal energy level structure as depicted
in Fig.~1(a). The atoms populate the state $\left|g\right\rangle$ and form a
Bose-Einstein condensate characterized by the field-operator $\widehat{\psi }_{g}
\left( {\bf r}\right)=\rho_{0}^{1/2}+\Delta\widehat{\psi }_{g}\left( {\bf r}
\right)$, where $\rho_{0}=N_{0}/V_{0}$ is the density of the (homogeneous)
condensate \cite{Footn:1} and the field operator $\Delta\widehat{\psi }_{g}\left(
{\bf r}\right)$ describes small deviations from an ideal condensate due to
non-condensate atoms. The Bose field-operators that account for the other two
electronic states are expanded in terms of plane waves as
\begin{equation} 
\widehat{\psi }_{j}\left( {\bf r}\right)
=V_{0}^{-1/2}\sum_{{\bf k}}b_{j,{\bf k}}e^{i{\bf k.r}},
\label{psi}
\end{equation}
where $b_{{\bf k},j}$ annihilates an atom with internal state $\left|j\right
\rangle$ ($j=q,e$) and wave-vector ${\bf k}$. A classical control field, with
wave-vector ${\bf k}_{c}$ and angular frequency $\omega_{c}$, couples states
$\left|q\right\rangle$ and $\left|e\right\rangle$. In the presence of this field,
the Hamiltonian for the excited atoms reads
\begin{eqnarray} 
H_{\rm atom}=\hbar\sum_{{\bf k}}\big[\omega_{e,{\bf k}}b_{e,{\bf k}}^{\dagger}
b_{e,{\bf k}}+(\omega_{q,{\bf k}-{\bf k}_c}+\omega_{c})b_{q,{\bf k}}^{\dagger }b_{q,{\bf k}}
\nonumber\\
+\Omega_{c}b_{q,{\bf k}-{\bf k}_c}^{\dagger }b_{e,{\bf k}}+\Omega_{c} 
b_{e,{\bf k}}^{\dagger }b_{q,{\bf k}-{\bf k}_c}\big],
\label{H-at}
\end{eqnarray} 
where $\omega_{j,{\bf k}}=\omega _{j}+\hbar k^{2}/ 2M\label{omega-j-k}$ ($j=q,e$)
are the atomic excitation energies (in units of $\hbar$), and $\Omega_{c}$ is the
Rabi frequency determining the magnitude of the excited-state splitting. To remove
an explicit time dependence due to the control field we have adopted a rotating
frame (frequency $\omega_{c}$) for the state $\left|q\right\rangle$. 

Adding now the interaction with a quantized probe field, the Hamiltonian for the
combined system of radiation and matter is written
\begin{equation} 
H =H_{\rm atom}+H_{\rm rad}+H_{\rm rad-atom}+H_{\rm cont}.
\label{H}
\end{equation}
Here $H_{\rm rad}$ is the radiative Hamiltonian,
\begin{equation} 
H_{\rm rad}=\sum_{{\bf k}}\hbar ck( a_{{\bf k}}^{\dagger }a_{{\bf k}}
+1/2),
\label{H-rad}
\end{equation}
where $a_{{\bf k}}$ is the annihilation operator for a probe photon
(polarized along the dipole moment ${\bf \mu}$ of the $g\rightarrow e$
transition), and the summation is over all wave-vectors, ${\bf k}\equiv{\bf k}_p$,
in the probe pulse wave-packet; the operator $H_{\rm rad-atom}=
-\varepsilon_{0}^{-1}\int {\bf d}({\bf r}){\bf p}({\bf r})d^{3}{\bf r}$
describes the interaction between the probe field and the atoms, where ${\bf d}
({\bf r})$ and  ${\bf p}({\bf r})={\bf \mu}\widehat{\psi }_{e}\left( {\bf r}
\right)^{\dagger}\widehat{\psi}_{g}\left({\bf r}\right)+{\rm h.c.}$ are the electric
displacement and polarization field operators, respectively; the last term,
$H_{\rm cont}=\left(x/2\varepsilon_{0}\right)\!\int{\bf p}({\bf r}){\bf p}
({\bf r})d^{3}{\bf r}$, represents the contact interaction which appears in
the multipolar formulation of QED \cite{Pow:64,Coh-Tan:89,Lew:94,Meystr:01},
commonly adopted with $x=1$. We choose $x=2/3$ in the contact interaction
\cite{Footn:2}. This yields the correct local field corrections in the refractive
index [Eq.~(\ref{n})].

As a first step we take into account only coherent optical processes in which
the absorption of a photon promotes a condensate atom to the excited electronic
state from which it returns, via photon emission, to the condensate. For this purpose,
we make the replacement $\widehat{\psi }_{g}\left( {\bf r}\right)\rightarrow
\rho_{0}^{1/2}$ in the polarization field ${\bf p}({\bf r})$.
In doing this, incoherent processes, such as spontaneous emission to modes other
than that of the absorbed photon, are disregarded. Spontaneous emission
will be analyzed in a second step later on. With the replacement, one has    
\begin{equation}
H_{\rm rad-atom}=-i\sum_{{\bf k}}
\mu\sqrt{\frac{\hbar ck\rho}{2\varepsilon_{0}}}
(a_{-{\bf k}}^{{}}-a_{{\bf k}}^{\dagger }) 
(b_{e,{\bf k}}+b_{e,-{\bf k}}^{\dagger })  
\label{H-rad-at}
\end{equation}
and
\begin{equation}
H_{\rm cont}=x\sum_{{\bf k}}\frac{\mu ^{2}\rho}{2\varepsilon_{0}}
(b_{e,-{\bf k}}+b_{e,{\bf k}}^{\dagger })
(b_{e,{\bf k}}+b_{e,-{\bf k}}^{\dagger }),
\label{H-cont}
\end{equation}
where the transition dipole moment is chosen to be real.

The complete Hamiltonian, with the simplified terms (\ref{H-rad-at}) and (\ref{H-cont}),
has been diagonalized by a Bogoliubov-type transformation, which mixes the radiation and
matter modes to yield the polaritons. We thus obtain, adapting the methods of 
Juzeli\={u}nas \cite{Juz:96},  
\begin{equation}
H=\hbar \sum_{{\bf k}}\sum_{m=1}^{3}\omega _{{\bf k}}^{(m)}P_{m,{\bf k}}^{\dagger}
P_{m,{\bf k}}+{\rm const},
\label{H-pol}
\end{equation}
where $m=1,2,3$ labels the polariton dispersion branches (Fig.~2) and the
eigen-frequencies $\omega _{{\bf k}}^{(m)}\equiv\omega $ are to be determined
from the equation $\omega=ck/n$, where $n\equiv n\left(\omega, {\bf k}\right)$ is
the refractive index, with
\begin{equation}
n^{2}=\frac{1+x\alpha\rho /\varepsilon _{0}}{1-\left(1-x\right)\alpha\rho
/\varepsilon _{0}} \qquad\left( x=2/3\right);
\label{n}
\end{equation}
$\alpha$ is the atomic polarizability, given for $\left|\Delta\omega\right|\ll\omega$ by 
\begin{equation}
\alpha=-\frac{\mu^{2}}{\hbar }\frac{\Delta\omega}{\Delta\omega ^2+\beta\Delta\omega-
\Omega_c^2}, 
\label{polariz}
\end{equation}
where $\Delta\omega=\omega -\omega_{c}-\omega_{q,{\bf k}-{\bf k}_c}$
is the detuning from the two-photon resonance and $\beta=\omega_{q,{\bf k}
-{\bf k}_c}+\omega_{c}-\omega_{e,{\bf k}}$ is the control laser frequency mismatch.

\begin{figure}[hb]
\vskip-0.25in 
\centerline{\hskip0.25725in\includegraphics[width=2.3625in,keepaspectratio]
{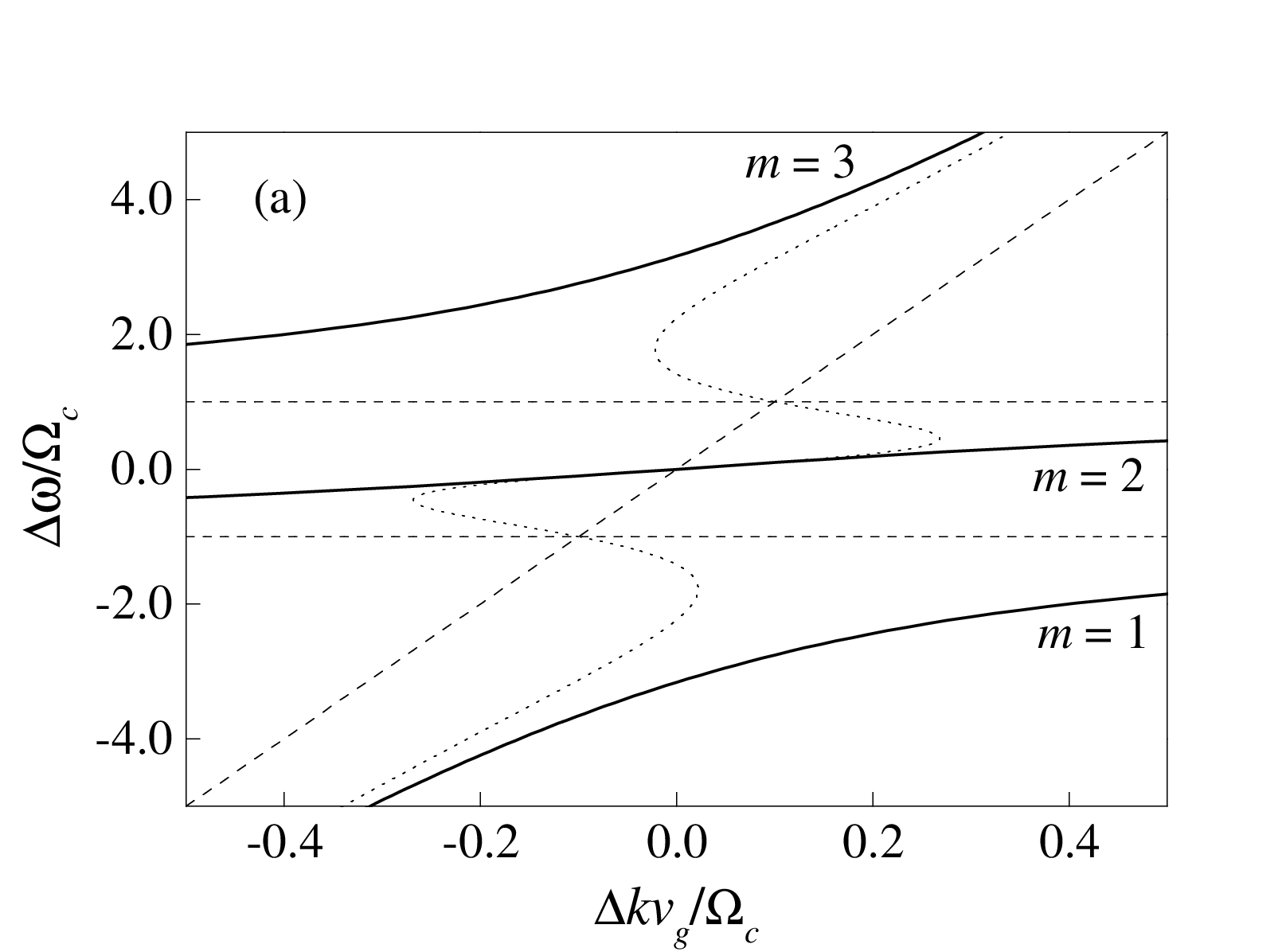}}
\vskip0.1in
\centerline{\includegraphics[width=2.2575in,keepaspectratio]
{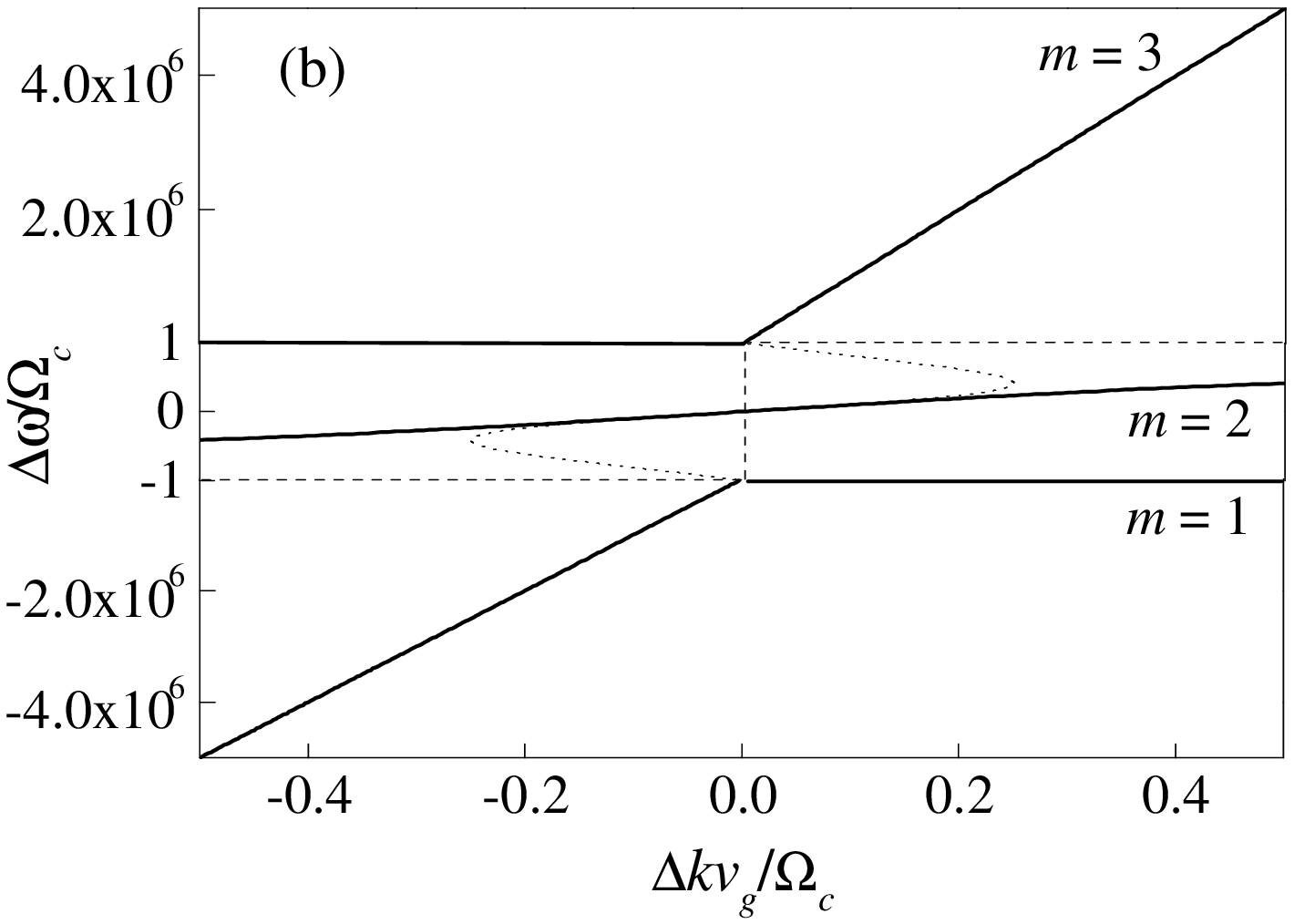}}
\vskip0.1in
\caption{Polariton dispersion branches (solid lines) in an EIT medium with $\beta=0$.
The dashed lines show the uncoupled modes in the limit of vanishing density. The
dotted curve is calculated from the semiclassical susceptibility and includes
the excited state linewidth. The parameters in (b) are taken from the experiment
of Liu {\it et al.} [6]; note the expanded frequency scale around the EIT resonance
where the free-space photon branch is almost a vertical line. In (a) the density is
reduced to give $v_g/c=10^{-1}$ so that the curves can be plotted using a
single frequency scale.
}
\end{figure}  

The polariton modes are defined by the Bose operators $P_{m,{\bf k}}$. These are
determined explicitly to be ($\left|\Delta\omega\right|\ll\omega$)
 \begin{equation}
P_{m,{\bf k}}=P_{m,{\bf k}}^{\rm (rad)}+iu  b_{q,{\bf k}-{\bf k}_c}+iu \left(\Delta
\omega /\Omega_c\right) b_{e,{\bf k}}, 
\label{P}
\end{equation}
where $u\equiv u_{m,{\bf k}}=\left(1 - n v_g/c \right)^{1/2}\![1+(\Delta\omega/
\Omega_c)^2]^{-1/2}$ and
\begin{equation}
P_{m,{\bf k}}^{\rm (rad)}=\left(v_g/c\right)^{1/2}[(n+1)a_{{\bf k}}+(n-1)
a_{{-\bf k}}^{\dagger }]/2
\label{P-rad}
\end{equation}
is the radiative component;
here $v_g=c\left(n+\omega\partial n/\partial\omega\right)^{-1}$ is the radiative
group velocity, and both $n\equiv n^{(m)}$ and $v_g\equiv v_g^{(m)}$ are to be
calculated at $\omega=\omega _{{\bf k}}^{(m)}$. 

In contrast to the polaritons considered in \cite{Mazets:96,Fleis:00,Fleis:01},
the operator  $P_{2,{\bf k}}$ represents an eigen-mode of the total Hamiltonian,
not only at EIT resonance ($\Delta\omega=0$), but for the entire branch of slow
polaritons ($m=2$). The systematic $\bf k$-space formulation we have presented
facilitates a discussion of the frequency and wave-vector dependence of the
radiation and matter fields.

The polariton dispersion branches are plotted in the vicinity of the EIT resonance
in Fig.~2. Figure 2(b), in particular, adopts the parameters of the recent experiment,
in which $v_g/c\approx2\Omega_{c}^{2}\hbar\varepsilon_{0}/\omega _{0}
\mu ^{2}\rho\approx10^{-7}$~\cite{Hau:01}. To aid comparisons with the EIT
literature, where the refractive index is usually plotted, we also plot the
dispersion curve (dotted line) given by the semiclassical susceptibility. The
susceptibility includes the effect of the excited state lifetime and hence the
dispersion curve continuously connects the three polariton branches.

In our theory the polaritons acquire a finite lifetime through the excited-state
contribution to Eq.~(\ref{P}). The decay rate is $\Gamma=u^2\left(\Delta\omega/
\Omega_c\right)^2\Gamma_0$, where $\Gamma_0$ is the rate of atomic radiative
decay and $u\approx 1$ provided $\left|\Delta\omega\right|\ll\Omega_c$. Away
from the EIT resonance, the polariton decays spontaneously at this rate into a
photon and a translationally excited ground-state atom. In the experiment
\cite{Hau:01} the probe pulse excites a frequency band $|\Delta\omega|/\Omega_c
\sim0.02$. There is thus significant loss ($\sim25\%$) due to polariton decay 
during the measured $11.8\,\mu{\rm s}$ propagation time in the medium. Some
broadening of the probe pulse is expected from this as the decay rate is
frequency dependent.

The contact interaction has no discernable effect on the slow polariton branch of
Fig.~2. At the density currently used \cite{Hau:01} its influence is at the level
of less than $1\%$. With a ten-fold increase in density, however, local field
effects would no longer be negligible. The distinction between the radiative and
full group velocity $\partial\omega/\partial k=\partial(\Delta\omega)/\partial k
+\partial\omega_{q,k-k_c}/\partial k\approx v_g+\hbar(k-k_c)/M$ is also unimportant
in current experiments.
The atomic velocity, $\hbar\left( k-k_c \right)/M$, is much smaller than $v_g$
even for counter-propagating control and probe lasers (typically it is of the
order of a few cm/s).

We consider now the time evolution of a wave-packet of EIT polaritons (a probe pulse in
the medium) influenced by the following sudden perturbations: (i) at $t=t_1$ the original
control beam is switched off, (ii) after some delay, at $t=t_2>t_1$, the control beam
is again turned on with a new Rabi frequency $\Omega_c^{\prime}$ and wavevector
${\bf k}_c^{\prime}$. In contrast to what has been previously considered
\cite{Hau:01,Phill:01,Fleis:00,Fleis:01}, ${\bf k}_c^{\prime}$ does not necessarily 
coincide with the original wavevector ${\bf k}_c$. If ${\bf k}_c^{\prime}=-{\bf k}_c$,
for instance, the regenerated probe pulse can move backwards, as we will show.
  
At $t<t_1$ the wave-packet is described by the state-vector  
\begin{equation} 
\left|t\right\rangle=\left|\{\alpha_{{\bf k},t}\}\right\rangle\equiv\prod_{{\bf k}}\exp\!
\big(\alpha_{{\bf k},t}P_{2,{\bf k}}^{\dagger }-{\rm h.c.}\big)
|{\rm vac}\rangle,  
\label{state-vector} 
\end{equation} 
where $\left|{\rm vac}\right\rangle$
is the vacuum state-vector ($P_{2,{\bf k}}\left|{\rm vac}\right\rangle =0$), $\left|
\{\alpha_{{\bf k},t}\}\right\rangle$ is a many-mode coherent state,  and the amplitude
$\alpha_{{\bf k},t}=\alpha_{{\bf k},t_1}\exp[-i\omega _{{\bf k}}^{(2)}(t-t_1)]$.
At $t=t_1$ the control laser is suddenly switched off. Equation (\ref{state-vector})
then represents the initial condition for the subsequent time evolution, giving for
$t_1<t<t_2$, 
\begin{equation} 
\left|t\right\rangle=\prod_{{\bf k}}\exp\!\big(iu \alpha_{{\bf k},t}
b_{q,{\bf k}-{\bf k}_c}^{\dagger }- {\rm h.c.}\big)\exp(S_{{\bf k}})
|{\rm vac}\rangle.   
\label{state-vector1} 
\end{equation} 
The first exponent
describes the magnetic excitations,
which are now decoupled from the photons so that (up to a phase) $\alpha_{{\bf k},t}
=\alpha_{{\bf k},t_1}\exp{[-i\omega _{q,{\bf k}-{\bf k}_c}(t-t_1)]}$. The operator
$\exp(S_{{\bf k}})$ accounts for the small radiative ($\sim3\times10^{-4}$) and excited
state ($\sim0.02$) contributions
to the slow polariton amplitude. These excitations are subsequently converted into
spontaneously emitted photons and may be omitted for $t-t_1$ larger than the radiative
lifetime. It should be emphasized that while the radiative contribution plays an
essential role in giving the polariton its velocity while the control laser is
``on,'' it accounts for a very small fraction of the quasiparticle number
(when $v_g/c\ll1$) and may simply be discarded to stop the polariton.

At $t=t_2$ the new control beam is applied. Expanding the magnetic operator
$ib_{q,{\bf k}-{\bf k}_c}^{\dagger }\equiv ib_{q,{\bf k}'-{\bf k}_c'}^\dagger$ in
terms of the new polariton operators $P_{m,{\bf k}'}'^\dagger$ ($m=1,2,3$),  
the state-vector evolves for $t>t_2$ as  
\begin{equation} 
\left|t\right\rangle=\prod_{{\bf k}'}\exp\!\big(uu'\alpha_{{\bf k}',t}^{\prime}
P_{2,{\bf k}'}'^\dagger-{\rm h.c.}\big)\exp(S_{{\bf k}'}^{\prime})|{\rm vac}\rangle,   
\label{state-vector2} 
\end{equation} 
with $\alpha_{{\bf k}',t}^{\prime}=\alpha_{{\bf k},t_2}\exp[-i\omega _{{\bf k}'}'^{(2)}
(t-t_2)]$. The operator $S_{{\bf k}'}$ accounts for the other polariton modes ($m=1,3$)
which are subsequently converted into spontaneously emitted photons. Here
${\bf k}^{\prime}={\bf k}-{\bf k}_c+{\bf k}_c^{\prime}$ is the wave-vector of the
regenerated polariton and the quantities $u'$, $P_{2,{\bf k}'}'$, and
$\omega_{{\bf k'}}'^{(m)}$ are defined in the obvious way \cite{Footn:2aa}.  

Such behavior may be viewed as a kind of time-delayed four-wave mixing (photon echo
\cite{Moss:79}) involving the original probe pulse, ${\bf k}$,
two control beams, ${\bf k}_c$ and ${\bf k}_{c}^\prime$, and the regenerated probe pulse,
${\bf k}^\prime$, the wave-vectors satisfying the phase matching condition ${\bf k}-
{\bf k}_c={\bf k}^\prime-{\bf k}_{c}^\prime$. The distinctive feature is the involvement
of slow EIT polaritons. Consequently very high conversion efficency can be reached ($u u'
\approx 1$) if both the original pulse and the regenerated polaritons are in the vicinity
of the EIT resonance. 
Note that there is no need for a smooth (adiabatic) ``turn-off'' and ``turn-on'' of the control
laser to map the slow polariton into and out of a magnetic excitation. Even if the
switching on and off are instantaneous, the reconstruction is almost perfect when $u u'\approx 1$.
The actual storage and retrieval of the probe pulse takes place at the medium boundaries
where vacuum photons are converted into slow polaritons and {\it vica versa}.    

We consider, finally, some specific situations. If the wavevector of the control laser
does not change (${\bf k}_{c}^\prime={\bf k}_c$), one arrives at degenerate four wave
mixing in which the regenerated probe photon has the same wavevector as the original
(${\bf k}^\prime={\bf k}$). This is the situation investigated in recent work
\cite{Hau:01,Phill:01,Fleis:00,Fleis:01}. On the other hand, with co-propagating control
and probe beams one can change the direction of the probe pulse by changing the
direction of the control beam \cite{Footn:3}. If ${\bf k}_{c}^\prime=-{\bf k}_c$
\cite{Footn:3a}, then ${\bf k}^\prime=-{\bf k}-2\left({\bf k}_c-{\bf k}\right)$, and
the reverted polariton experiences a frequency shift $\delta\omega^\prime\approx2v_{g}'
(k_c-k)$. Under the conditions of the recent experiment \cite{Hau:01}
one obtains $\delta\omega^\prime\approx0.26\,{\rm kHz}\times2\pi$, which is very small
compared to the coupling Rabi frequency (and also the spectral width of the probe pulse)
and the reverted polariton remains in the EIT region.
For counter-propagating beams, the wave-vector of the new polariton is
${\bf k}^\prime={\bf k}+2{\bf k}^\prime_c\approx 3{\bf k}$; the polariton becomes
situated close to the upper band edge where there is a rate of radiative decay on the
order of $\Gamma_0$ (unless $\Omega_c^\prime$ is extremely small) \cite{Footn:4}.

We have formulated a theory of slow polaritons in atomic gases and applied
it to the slowing down, storing, and redirecting of laser pulses in an EIT medium. The
polariton modes have been determined through a full diagonalization of the dissipationless
Hamiltonian, and lifetimes introduced as a secondary step. With detuning included various
four-wave mixing possibilities were analyzed. The possibility of reverting a stopped
polariton was demonstrated. 

The authors acknowledge helpful discussions with T.~W.~Mossberg, M.~G.~Raymer,
M.~Ma\v{s}alas, and K.~V.~Krutitsky.   
Work supported by the NSF under Grant No.\ PHY-0099576 and by the Council
for International Exchange of Scholars.


\begin{references}
\bibitem{Arimondo:96} E. Arimondo, In {\it Progress in Optics}, edited by E. Wolf\
(Elsevier, Amsterdam, 1996), p.257.
\bibitem{Harris:97} S.E. Harris, Physics Today {\bf 50}(7), 36 (1997).
\bibitem{Scully:book} M.O.~Scully and M.S.~Zubairy, {\it Quantum Optics\/}
(Cambridge University Press, Cambridge, 1997).
\bibitem{Bergman:98} K.~Bergmann, T.~Theuer and B.W.~Shore, Rev.\ Mod.\ Phys.\
{\bf 70}, 1003 (1998). 
\bibitem{Hau:99} L.V.~Hau, S.E.~Harris, Z. Dutton, and C.H.~Behroozi, Nature 
{\bf 397}, 594 (1999). 
\bibitem{Hau:01} C. Liu, Z. Dutton, C.H.~Behroozi and L.V.~Hau, Nature 
{\bf 409}, 490 (2001). 
\bibitem{Phill:01} D.F.~Phillips, A. Fleischhauer, A. Mair, R.L.~Walsworth and
M.D.~Lukin, Phys.\ Rev.\ Lett.\ {\bf 86}, 783 (2001).
\bibitem{Juz:96}  G. Juzeli\={u}nas, Phys.\ Rev.\ {\bf A 53}, 3543 (1996).
\bibitem{Knoe:89}  J. Knoester, and S. Mukamel, Phys.\ Rev.\ A {\bf 40}, 7065
(1989).
\bibitem{Svist:90}  B. Svistunov and G. Schlyapnikov, Zh.\ Eksp.\ Theor.\
Phys.\ {\bf 98}, 129 (1990) [Sov.\ Phys.\ JETP {\bf 71}, 71 (1990)].
\bibitem{Politz:91}  H.D.~Politzer, Phys.\ Rev.\ A {\bf 43}, 6444 (1991).
\bibitem{Barn:92} B. Huttner and S.M.~Barnett, Phys.\ Rev.\ A {\bf 46}, 4306
(1992); S.-T.~Ho and P. Kumar, J.\ Opt.\ Soc.\ Am.\ B {\bf 10}, 1620
(1993).
\bibitem{Drum:99} P.D.~Drummond and M. Hillery, Phys.\ Rev.\ {\bf A 59}, 691
(1999).
\bibitem{Juz:00}  G. Juzeli\={u}nas and D.L.~Andrews, In {\it Advances in Chemical Physics},\ 
v.112, edited by I. Prigogine and S. A. Rice (Wiley, New York, 2000), p.357.
\bibitem{Mazets:96} I. E. Mazets and B. G. Matisov, JETP Lett. {\bf 64},
515 (1996) [Pis'ma\ Zh.\ Exp.\ Theor.\ Fiz. {\bf 64}, 473 (1996)].
\bibitem{Fleis:00} M. Fleischhauer and M.D.~Lukin, Phys.\ Rev.\ Lett.\ {\bf 84},
5094 (2000).
\bibitem{Fleis:01} M. Fleischhauer and M.D.~Lukin, quant-ph/0106066.
\bibitem{Footn:1} The homogeneous model may also be suitable for the
analysis of a trapped BEC in the local density (Thomas-Fermi)
approximation, see G. Morigi and G.S.~Agarwal, Phys.\ Rev.\ {\bf A 62},
013801 (2000).
\bibitem{Pow:64} E.A.~Power, {\it Introductory Quantum Electrodynamics\/}
(Longmans, London, 1964).
\bibitem{Coh-Tan:89} C.~Cohen-Tannoudji, J.~Dupont-Roc, and G.~Grynberg,
{\it Photons and Atoms\/} (Wiley, New York, 1989).   
\bibitem{Lew:94}  M.~Lewenstein, L.~You, J.~Cooper and K.~Burnett, Phys.\
Rev.\ A {\bf 50}, 2207 (1994).
\bibitem{Meystr:01} P.~Meystre, {\it Atom Optics\/} (Springer, Berlin, 2001).  
\bibitem{Footn:2} In fact, by taking $x=1$, one implicitly assumes that the
electrostatic interaction between atoms is given by $\left( 2\varepsilon_{0}
\right)^{-1}\!\int{\bf p}^{\|}({\bf r}){\bf p}^{\|}({\bf r})d^{3}{\bf r}$,
which contains the Dirac delta singularity at the origin characteristic of
the dipole-dipole coupling between point atoms. This is an
artifact because atoms, which are not point objects,  do not overlap
in reality. The singularity is eliminated by subtracting the term
$\left(6\varepsilon_{0}\right)^{-1}\!\int{\bf p}({\bf r}){\bf p}({\bf r})
d^{3}{\bf r}$, leading to our $x=2/3$.
\bibitem{Footn:2aa} The frequency $\omega_c^\prime$ might be tuned to another excited
level $e^\prime$ leading to a substantial change in the frequency of the regenerated
polariton.  
\bibitem{Moss:79}  T.W.~Mossberg, R.~Kachru, S.R.~Hartmann and A.M.~Flusberg,
Phys.\ Rev.\ A {\bf 20}, 1976 (1979).
\bibitem{Footn:3} This reversion of the probe pulse differs from that achieved
by making the group velocity negative in a Doppler broadened gas: O.~Kocharovskaya,
Y.~Rostovtsev and M.O.~Scully, Phys.\ Rev.\ Lett.\ {\bf 86}, 628 (2001).
\bibitem{Footn:3a}  Under the conditions of the experiment \cite{Hau:01}, the
circular polarization of the control laser should be reversed in order to drive the same
transition $q\rightarrow e$.
\bibitem{Footn:4} Such a polariton cannot leave the sample because it is far
from the crossing point with the photon dispersion. Another reversion of the
control laser would allow it to escape.
\end{references}
\end{document}